\documentclass[12pt]{article}
\usepackage[utf8x]{inputenc}
\usepackage[T1]{fontenc}
\usepackage{amssymb,amsmath,mathtools,amsfonts}
\usepackage{graphicx}
\usepackage{lmodern}
\usepackage{float}
\usepackage[colorlinks=true, allcolors=blue]{hyperref}
\usepackage{tabularx,booktabs}
\usepackage[superscript]{cite}
\usepackage{abstract}
\usepackage{caption}
\usepackage{titlesec}
\usepackage{afterpage}
\usepackage{ragged2e} 

\evensidemargin\oddsidemargin \hoffset-22.4mm \voffset-28.4mm
\graphicspath{{./fig/}{./fig_SuppPub/}}

\begin{document}

\title{Oscillation threshold of a Raman clarinet with localized nonlinear losses at the open end}
\author{Nathan Szwarcberg \textsuperscript{a, b}, Tom Colinot \textsuperscript{a, b}, Christophe Vergez \textsuperscript{b}, Michaël Jousserand \textsuperscript{a} \\
\textsuperscript{a} Buffet Crampon, 5 Rue Maurice Berteaux, 78711 Mantes-la-Ville, France\\
\textsuperscript{b} Aix Marseille Univ, CNRS, Centrale Med, LMA, Marseille, France
}
\date{Submitted 2024/10.}
%\twocolumn[
\maketitle
\begin{onecolabstract}
Localized nonlinear losses are taken into account in a simple Raman clarinet model.
The complete system is expressed as an iterated map, enabling to study the stability of the different playing regimes. 
A parametric study is carried out with respect to three major parameters: blowing pressure, embouchure and nonlinear losses coefficient.
The model exhibits the well-known effect of {reducing the maximum blowing pressure until the oscillations stop (extinction threshold)} when nonlinear losses increase.
{Furthermore}, the stability analysis also shows that increasing nonlinear losses increases the {minimal blowing pressure for which the oscillations start (oscillation threshold)}.
  \end{onecolabstract}
  \vspace{10pt}
%]
\maketitle
\section{Introduction}
	The Raman model can be considered as {one of} the simplest physical models of clarinet\cite{bible2016}. 
	It is composed of a reed which behaves as an idealized spring with no other dynamics, and of a passive resonator characterized by its reflection function.
	Viscothermal losses are neglected or reduced to a constant coefficient\cite{maganza1986bifurcations} independent from the frequency.
	The motion of the reed is coupled to the response of the resonator through the nonlinear characteristic of the flow crossing the reed channel\cite{wilson_operating_1974}, enabling the system to produce self-sustained oscillations.

	In spite of its simplicity, the Raman clarinet model {is useful in that in can provide valuable insights on the behavior of} a real instrument.
	{The model predicts experimental results} on the position and nature of the oscillation and extinction thresholds\cite{dalmont_oscillation_2007, dalmont2005analytical, atig2004saturation}, which are the minimal and maximal blowing pressures enabling self-sustained oscillations.
	Moreover, when the blowing pressure is increased linearly in time, experimental and numerical results both highlight a delay in the onset of the oscillation threshold\cite{bergeot2014response}  . This delay indicates a sensitivity of the system to dynamic bifurcations\cite{bergeot2013prediction}.
Finally, the influence of localized nonlinear losses
{at the end of the resonator} on the extinction threshold has been demonstrated by a comparison between experiments and simulations with a modified Raman model\cite{dalmont_oscillation_2007,  atig2004saturation}.

Although the influence of localized nonlinear losses on the dynamic behavior of the clarinet is documented for very high blowing pressure\cite{dalmont_oscillation_2007,  atig2004saturation,szwarcberg2023amplitude}, their influence over the global range of control parameters is less well known.
Keefe (1983)\cite{keefe1983acoustic} points out that it is more difficult for a clarinetist to maintain a stable oscillating regime when the contribution of localized nonlinear losses in a tone hole is higher.
This observation invites us to study the {dynamic behavior} of a simple modified Raman model {made of a cylindrical tube with localized nonlinear losses at the open end}.

{Stability and multistability in the system determine the consistency and predictability of the produced sound. 
A stable oscillating regime corresponds to an easily controllable tone, while an unstable regime cannot sound for more than a fleeting second, unless a specific technique is employed to maintain it. 
Multistability occurs when different regimes coexist for the same control parameters\cite{colinot2021multistability}.
In this case, a musician may need to apply precise initial conditions to play the desired regime. 
Additionally, the nature of the bifurcations affects how smoothly the transition from a regime to another occurs. 
In a direct Hopf bifurcation, the sound emerges continuously as the blowing pressure increases, facilitating the production of a \textit{pianissimo}. 
In contrast, an inverse Hopf bifurcation leads to an abrupt emergence of the oscillations, making it harder for the musician to control soft dynamics\cite{taillard2015analytical}.
Understanding how localized nonlinear losses affect the stability of the oscillating regimes and the type of the bifurcations could help inform decisions in instrument design, as well as providing insights for musicians.
}

The complete model accounting for nonlinear losses located at the open end of the tube is defined in section \ref{sec:model}.
The system is studied as an iterated map\cite{maganza1986bifurcations, taillard2010iterated}.
Thus, the stability of the different regimes can be easily computed with respect to the blowing pressure, the embouchure parameter and the nonlinear losses coefficient.
The method to study the stability of the different regimes is recalled in section \ref{sec:methods}.
The influence of localized nonlinear losses on the stability of the equilibrium and on the two-state regime is studied in section \ref{sec:results}.
Other {long-period} regimes are also investigated.
Finally, in section \ref{sec:discussion}, the influence of localized nonlinear losses on the {stability of the oscillating regimes and the nature of the bifurcations} is compared with that of linear losses.
Possible implications for the musician are discussed.

\section{Model} \label{sec:model}
A {cylindrical} tube is considered of length $L$, radius $a$ and cross-section $S$.
The characteristic impedance of plane waves in the tube is $Z_c=\rho_0 c_0/S$, where $\rho_0=1.23$~kg$\cdot$m$^{-3}$ is the density of air, and $c_0=343$~m$\cdot$s$^{-1}$ is the speed of sound in the air.
The position along the main axis of the tube is indicated by $x\in [0,L]$.

\subsection{Resonator}
The acoustic pressure $p(x,t)$ is defined as the sum of a {traveling} wave $p^+(x,t)$ and a regressive wave $p^-(x,t)$:
\begin{equation}\label{eq:p_p+}
p(x,t) = p^+(x,t)+p^-(x,t).
\end{equation}
The acoustic velocity $v(x,t)$ is defined from $p^+$ and $p^-$ as
\begin{equation}\label{eq:v_p+}
\rho_0 c_0 v(x,t) = p^+(x,t) -p^-(x,t).
\end{equation}
The acoustic flow $u$ is also defined so that $u=Sv$.

In the Raman model, viscothermal losses {do not depend on} the frequency\cite{maganza1986bifurcations}. 
Thus in the time domain, {the propagating waves are related as follows:}
\begin{equation}\label{eq:transport}
\left\{
\begin{matrix}
p^+(L,t) = \lambda p^+(0,t-L/c_0)\\
\lambda p^-(L,t- L/c_0) = p^-(0,t),
\end{matrix} \right.
\end{equation}
where $\lambda \in [0,1]$.
{It can be shown that in the Raman model, increasing the dissipation due to radiation at the open end of the tube is equivalent to reducing $\lambda$: losses due to propagation and localized linear losses due to radiation are indistinguishable.}

\subsection{{Boundary condition for the localized nonlinear losses}}
In this section, for reading comfort, $p^+(L,t)$ and $p^-(L,t)$ are written as $p^+$ and $p^-$.
The same applies for $p$ and $v$.

{Localized nonlinear losses are related to the dissipation of the turbulent flow at the exit of the tube, which can take the form of vortices for high acoustic velocities\cite{ingaard1950acoustic}.
They are accounted for through the following relationship at $x=L$, assuming a quasi-stationary flow}\cite{disselhorst1980flow,singh2014nonlinear,atig2004saturation}:
\begin{equation}\label{eq:bc}
 p =\rho_0 C_\mathrm{nl}\,  v|v|,
\end{equation}
where $C_\mathrm{nl}\geq 0$ is a dimensionless constant {which characterizes the vena-contracta of the jet}.
{Its value increases when the sharpness of the edges at the open end of the tube increases.  
          Moreover, it does not depend on the diameter of the tube \cite{komkin2020experimental}.   
     Theoretically, the maximum value of the nonlinear losses coefficient is $C_\mathrm{nl}=0.5$ (often written $c_d=4\cdot C_\mathrm{nl}=2$) for a monochromatic wave. 
     This theoretical boundary applies to an unflanged tube with thin walls and sharp edges \cite{peters1993damping}.
     However, Atig \textit{et al.\ }(2004)\cite{atig2004saturation} and Dalmont and Frapp\'e (2007) \cite{dalmont_oscillation_2007} find a best fit on experimental results for $C_\mathrm{nl}=0.7$ ($c_d=2.8$), for a clarinet with an unflanged termination with sharp edges.
     As explained by Dalmont and Frapp\'e (2007) \cite{dalmont_oscillation_2007}, this overestimation can first be due to the nature of the acoustic field in the clarinet which is significantly different from a sine wave, and also to the fact that in the Raman model, linear losses are underestimated for higher frequencies.
     }
     
{To suit the reflection function formalism,}  the boundary condition \eqref{eq:bc} has to be rewritten by an explicit relationship between $p^-$ and $p^+$.
An elegant method to solve this problem is proposed by Monteghetti (2018, Chap. 2.4.2)\cite{monteghetti2018analysis}.
In the present article, {another} and more straightforward method is employed by substituting directly $p^+$ and $p^-$ in Eq. \eqref{eq:bc} through Eqs. \eqref{eq:p_p+} and \eqref{eq:v_p+}.
Two quadratic polynomials in $p^-$ are obtained after this substitution:
\begin{align*}
(p^-)^2 - \left(2 p^+ + K^{-1} \right)p^- + \left( p^+ - K^{-1} \right) p^+ &= 0 ~ \mathrm{if}~p^+\geq p^-, \\ 
(p^-)^2 - \left(2 p^+ - K^{-1} \right)p^- + \left( p^+ + K^{-1} \right) p^+ &= 0 ~ \mathrm{if}~p^+\leq p^-,
\end{align*}
where $K=C_\mathrm{nl}/(\rho_0 c_0^2)$.
The solution of each of these two equations in $p^-$ is:
\begin{align*}
p^-&= p^+ + \dfrac{1}{2K} \left( 1 -  \sqrt{1+ 8K p^+ } \right) \quad \mathrm{if~}p^+\geq0, \\
p^-&= p^+ + \dfrac{1}{2K} \left( -1 +  \sqrt{1- 8K p^+ } \right) \quad \mathrm{if~}p^+\leq 0.
\end{align*}
By combining the two conditions on $p^+$, the following expression is obtained for the reflection at the open end:
\begin{equation*}
p^- = p^+ +\dfrac{1}{2 K} \frac{p^+}{|p^+|} \left( 1 -  \sqrt{1+ 8K |p^+| }  \right).
\end{equation*} 
After a bit of algebra, the following expression is finally obtained: 
\begin{equation}\label{eq:refL2}
 p^-= p^+ \left(1- \frac{4}{1+ \sqrt{1+K_L|p^+|}} \right),
\end{equation}
with $K_L= 8 C_\mathrm{nl}/(\rho_0 c_0^2)$.
The reflection function at $x=0$ is deduced by substitution of Eq. \eqref{eq:transport} in Eq. \eqref{eq:refL2}:
\begin{align}
&p^-(0,t) = r_\mathrm{nl}[p^+(0,t-\tau)], \qquad \mathrm{where}\\
&r_\mathrm{nl}(\xi) = \lambda^2 \xi \cdot \left(1- \frac{4}{1+ \sqrt{1+ K_0 |\xi|}} \right),\label{eq:ref0}
\end{align}
with $K_0=\lambda K_L$ and $\tau=2L/c_0$.
It can be checked that the reflection function is passive, i.e for all $t>0$ and $\xi\in \mathbb{R}$: 
$$\int_{-\infty}^t |[r_\mathrm{nl}(\xi)](t)|^2\, \mathrm{d}t \leq \int_{-\infty}^t|\xi(t)|^2 \, \mathrm{d}t.$$

\begin{figure}
	\centering
	\includegraphics[width=.6\textwidth]{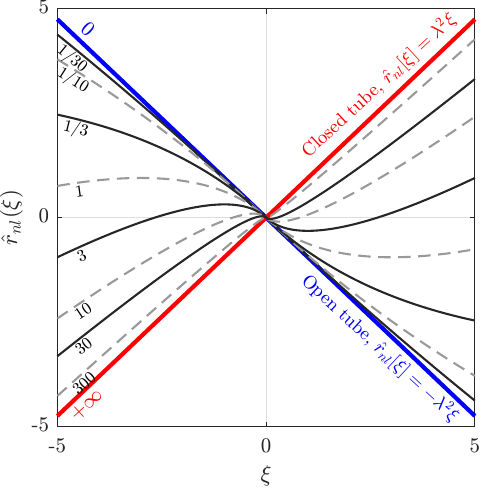}
	\caption{Evolution of the shape of $\hat{r}_\mathrm{nl}$ when increasing $\hat{K}_{0}$ (values are displayed on the figure), with $\lambda= \sqrt{0.95}$.}
	\label{fig:rnl}
\end{figure}

The reflection function (under its dimensionless form) is represented on Figure \ref{fig:rnl} for different values of ${K}_{0}\in [0, \infty]$.
Increasing nonlinear losses results in a continuous transformation from the boundary condition of an open tube (in blue) to that of a closed tube (in red).
This transformation is symmetrical with respect to the point $(0,0)$.

Note that the computation of the reflection function leads to a different result from Eq.~(22) of Atig \textit{et al.\ }(2004)\cite{atig2004saturation}.
{It can be checked that the expression from Atig \textit{et al.\ }(2004)\cite{atig2004saturation} is derived from Eq.\ \eqref{eq:refL2} by expanding $p^+$ to the first order around zero.}

\subsection{Nonlinear characteristic and complete model}
The flow entering the bore {through} the reed channel is given by the well-known nonlinear characteristic\cite{taillard2010iterated}:
\begin{equation}\label{eq:caractNL}
\hat{u}(t) = \zeta [\hat{p}(t)- \gamma +1]^+ \,  \text{sgn}[\gamma - \hat{p}(t)] \, \sqrt{|\gamma- \hat{p}(t)|},
\end{equation}
where $[\xi]^+=\xi\cdot\mathrm{Heaviside}(\xi)$, $\gamma>0$ is the dimensionless blowing pressure, and $\zeta\in [0,1]$\cite{taillard2010iterated} is the embouchure parameter.
The dimensionless input pressure and flow are noted $\hat{p}(t)= p(0,t)/P_M$ and $\hat{u}(t)=u(0,t) Z_c/P_M$, where {$P_M \in [4, 8.5]$~kPa}\cite{atig2004saturation, dalmont_oscillation_2007} is the minimal reed closing pressure.

The expression of the nonlinear characteristic \eqref{eq:caractNL} as an explicit relationship between  $p^+$ and $p^-$ can be found in Taillard \textit{et al.\ }(2010, Appendix A)\cite{taillard2010iterated} by setting: 
\begin{align*}
X &= \gamma - \hat{p} = \gamma - \hat{p}^+ - \hat{p}^-  \\
Y &= \hat{u} + X = \gamma -2 \hat{p}^-.
\end{align*}
The nonlinear characteristic can be written as follows:
\begin{align*}
&Y=X \qquad \mathrm{if}~X>1; \\
&Y = X + \zeta (1-X) \sqrt{X}  \qquad \mathrm{if}~0<X<1;\\
&Y = X - \zeta (1-X) \sqrt{-X} \qquad \mathrm{if}~X<0.
\end{align*}
The inverse function $X(Y)=Y^{-1}(X)$ is determined by the same authors.
From the definition of $X$, one can write:
\begin{align*}
\hat{p}^+ (t)=\gamma-X(Y)-\hat{p}^-(t).
\end{align*}
By substitution of $r_\mathrm{nl}$ [Eq. \eqref{eq:ref0}] in $\hat{p}^-(t)$, the complete system is summarized in the following expression:
\begin{align}\label{eq:iterate}
&\hat{p}^+ (t) = f\left(\hat{p}^+ (t-\tau)\right), \quad \mathrm{where} \\
&f(\xi)=\gamma-X\left(\gamma - 2 \hat{r}_\mathrm{nl}\left(\xi \right) \right)-\hat{r}_\mathrm{nl}\left( \xi \right) \quad \mathrm{and}\label{eq:ref0Adim}\\
&\hat{r}_\mathrm{nl}(\xi) = \lambda^2 \xi \cdot \left(1- \frac{4}{1+ \sqrt{1+ \hat{K}_0 |\xi|}} \right),
\end{align}
with $\hat{K}_0=P_M K_0$.
These formulas could be used in a delay-line formalism similar to Guillemain and Terroir (2006)\cite{guillemain2006digital}.
In the following, Eq.~\eqref{eq:iterate} is written as an iterated function: $\hat{p}^+_{n+1}=f(\hat{p}^+_{n})$, with a time increment of $\tau$.
In the following, for conciseness, $\hat{p}_n^+$ is denoted as $x_n$ so that $f(x_n)=x_{n+1}.$ 

\begin{figure}
	\centering
	\includegraphics[width=.6\textwidth]{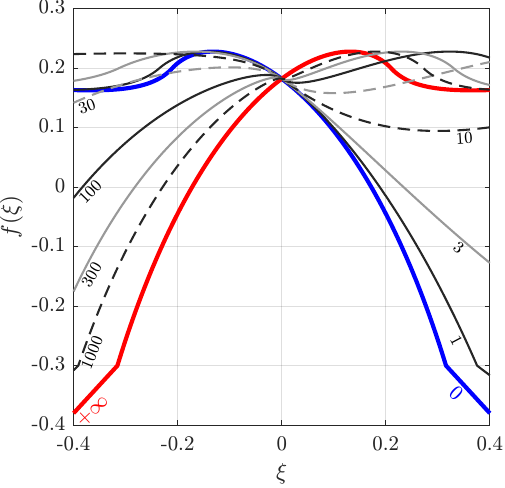}
	\caption{Evolution of $f$ when increasing $\hat{K}_0$ (values are displayed on the figure), with $\lambda=\sqrt{0.95}, \gamma=0.4$ and $\zeta=0.5$.}
	\label{fig:f}
\end{figure}

The {iteration} function is represented on Figure \ref{fig:f}, for $\lambda=\sqrt{0.95}, \gamma=0.4$, $\zeta=0.5$ and increasing values of $\hat{K}_{0}$ from the open tube ($\hat{K}_{0}=0$) to the closed tube ($\hat{K}_{0}\to +\infty$) boundary conditions.
All curves cross the point $(0,f(0))$ since $\hat{r}_\mathrm{nl}(0)=0$.
Moreover, the curves for $\hat{K}_{0}=0$ and $\hat{K}_{0}=+\infty$ are symmetrical with respect to the axis $\xi=0$, since $\hat{r}_\mathrm{nl}(\xi)$ for $\hat{K}_{0}=0$ is equal to $\hat{r}_\mathrm{nl}(-\xi)$ for $\hat{K}_{0}=+\infty$.

{According to the experimental results from Atig \textit{et al.\ }(2004)\cite{atig2004saturation} and Dalmont and Frapp\'e (2007)\cite{dalmont_oscillation_2007}, the maximum value of the localized nonlinear losses coefficient would be $\hat K_0=0.325$, for $C_\mathrm{nl}=0.7,  P_M=8.5~$kPa and $\lambda=\sqrt{0.95}$.
The motivation to study values of $\hat K_0$ that are far higher than observed experimental results is in line with the very simple formalism of the iterated functions, which calls for an exhaustive study of the space of the control parameters.
Furthermore, exploring the behavior of the clarinet for very high values of $\hat K_0$ is meaningful when nonlinear losses are localized in the side holes of a woodwind. 
As stated in Chapter 8.4.5.2 of Chaigne and Kergomard (2016)\cite{bible2016}: ``if the, even real, impedance of a side hole, increases notably, it may eventually act as a closed
end for the hole!''
}

\section{Methods}\label{sec:methods}
The properties of the {iteration} function are thoroughly described by Taillard \textit{et al.\ }(2010)\cite{taillard2010iterated}.
A few notions are recalled here to study the stability of the different regimes.

Function $f$ is an iterated function, meaning that
\begin{align*}
 x_{n+k}=\underbrace{\left(f \circ f \circ ... \circ f \right)}_{k\in \mathbb{N}^*} (x_n)= f^{(k)}(x_n). 
\end{align*}
For a given set of control parameters ($\gamma, \zeta, \hat{K}_0, \lambda$), a fixed point $x^*$ of $f$ is considered, so that 
$$f(x^*)=x^*.$$
From the practical point of view, {$f$ is sampled over a vector $\mathbf x$ with $N_x$ values $x_i$ linearly spaced within the range $x_i \in [-\gamma-0.1, \gamma+0.1]$ (the boundaries are chosen empirically to ensure that all fixed points are found for all values of the control parameters considered)}. 
{The fixed point} $x^*$ is computed by linear interpolation of the zero-crossing of {$f(\mathbf x)- \mathbf x$. The choice of linear interpolation is preferred since the iterates $f^{(k)}$ have many discontinuities on their derivatives.}
The fixed point is stable if $$|f'(x^*)|<1,$$ where $f'=\partial_x f$.
Furthermore, if $x^*_k$ is a fixed point of $f^{(k)}$, then it is also a fixed point for every iterate $f^{(nk)}$ with $n\in \mathbb{N}^*$. 
If $x^*_k$ is stable (resp.\ unstable) for the iterate $f^{(k)}$, then it is also stable (resp.\ unstable) for every higher iterates $f^{(nk)}$.
In this study, the stability of the following regimes is considered.
\begin{enumerate}
%	\item The equilibrium (non-oscillating state) denoted R$_1$, which is stable if one of the fixed points $x^*_1$ of $f$ is stable.
\item The equilibrium (non-oscillating state, denoted R$_1$) associated to a fixed point $x_1^*$ of $f$, which is stable if $x_1^*$ is stable with respect to $f$.
	\item The two-state (or quarter-wavelength) regime, denoted R$_2$,	which is stable if two fixed points $x_2^*$ of $f^{(2)}$ that are different from $x_1^*$ are stable with respect to $f^{(2)}$.
	{An example is shown on the first row of Suppl.\ Fig.\ 1.}
	\item {The long-period} regime (R$_n$), which is stable if $n$ fixed points $x_n^*$ of $f^{(n)}$ are stable with respect to $f^{(n)}$ and are different from the fixed points of $f^{(k)}$, where $k$ are divisors of $n$.
	In this article, only the regimes R$_3$, R$_4$, R$_6$ and R$_8$ are studied.
	{Examples of three-state and four-state regimes are provided on the second and third row of Suppl.\ Fig.\ 1.}
\end{enumerate}
To compute the stability of the different regimes of interest, the control parameters space $(\gamma, \zeta, \hat{K}_0, \lambda)$ is {meshed on a regular grid. 
The ranges of the parameters and their associated step sizes are given in Table \ref{tab:1}.}
{For the simulations leading to the results presented in sections \ref{sec:results} and \ref{sec:discussion_nl}}, the linear losses parameter is fixed at $\lambda=\sqrt{0.95}$ to reduce the problem to three dimensions and simplify the representation of the results. 
The value of $\lambda$ is chosen to be consistent with Taillard \textit{et al.\ }(2010)\cite{taillard2010iterated}.
For a detailed study about the sensitivity of the Raman model to $\lambda$, see Taillard and Kergomard (2015)\cite{taillard2015analytical} {and section \ref{sec:discussion_lin}}.

For every set of control parameters, the fixed points of $f, f^{(2)}$ and $f^{(n)}$ are computed.
The stability of the corresponding regimes is then determined.

\begin{table}[h!]
\centering
\caption{Ranges and step sizes of the control parameters used in the simulations.}
\label{tab:1}
\begin{tabular}{cccc} 
\hline
 Parameter & Min & Max & Step size \\
 \hline
 $\gamma$ & 0 & 5 & $10^{-3}$\\
 $\zeta$ & 0.01 & 0.99 & $5 \cdot 10^{-3}$\\
 $\lambda$ & 0.1 & 1 & $5 \cdot 10^{-3}$\\
 $\hat K_0$ & 0 & 100 & $5\cdot 10^{-2}$
\\
\hline 
\end{tabular}
\end{table}
%\newpage

\afterpage{
    \clearpage
\begin{figure}[p]
	 	\centering
	 	\includegraphics[width=.42\textwidth]{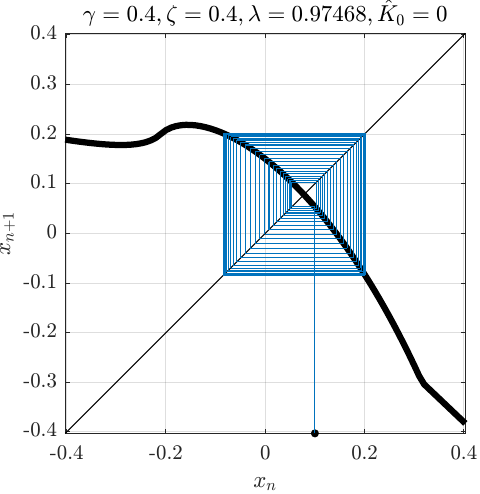}
	 	\hfill
	 	\includegraphics[width=.53\textwidth]{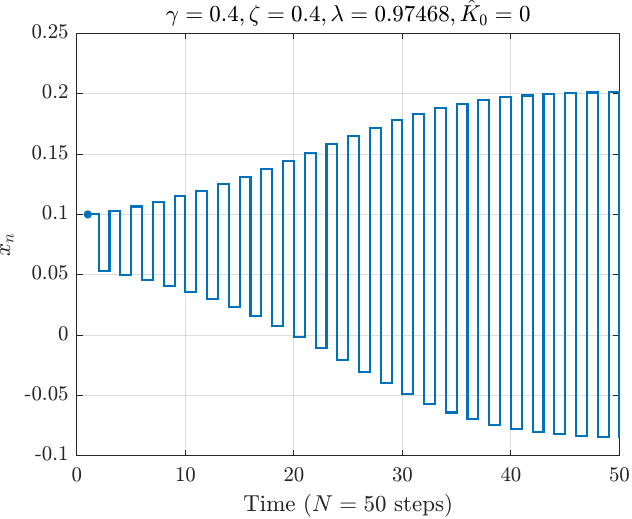}
	 	\includegraphics[width=.42\textwidth]{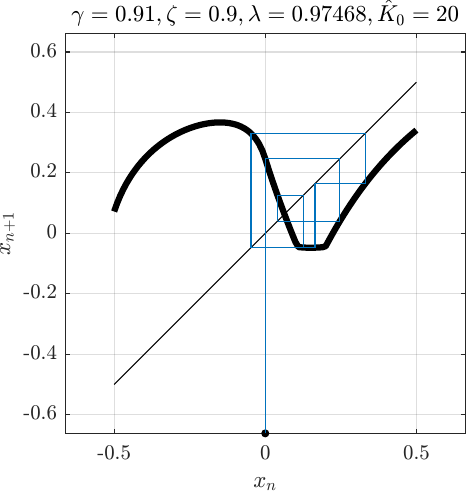}
	 	\hfill
	 	\includegraphics[width=.53\textwidth]{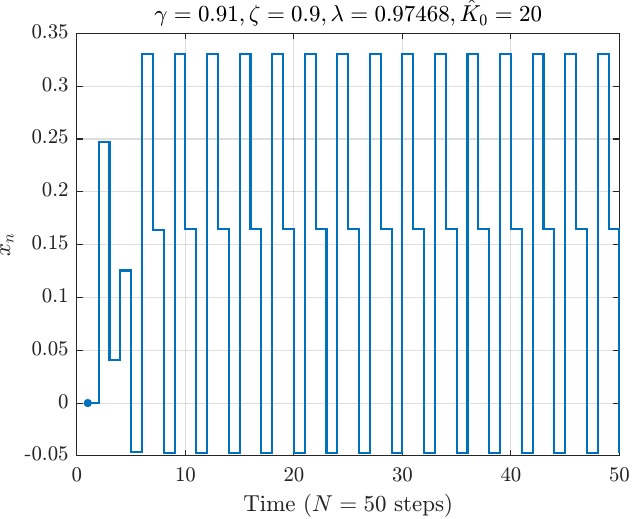}
	 	\includegraphics[width=.42\textwidth]{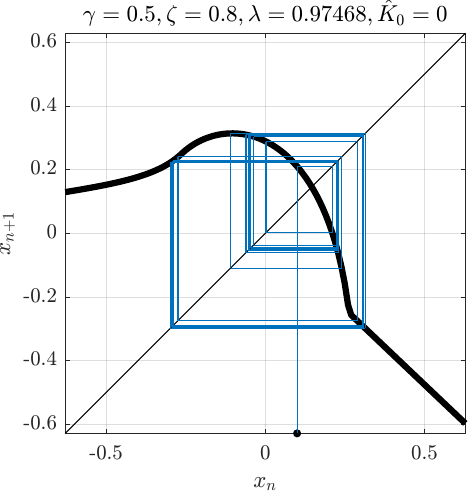}
	 	\hfill
	 	\includegraphics[width=.53\textwidth]{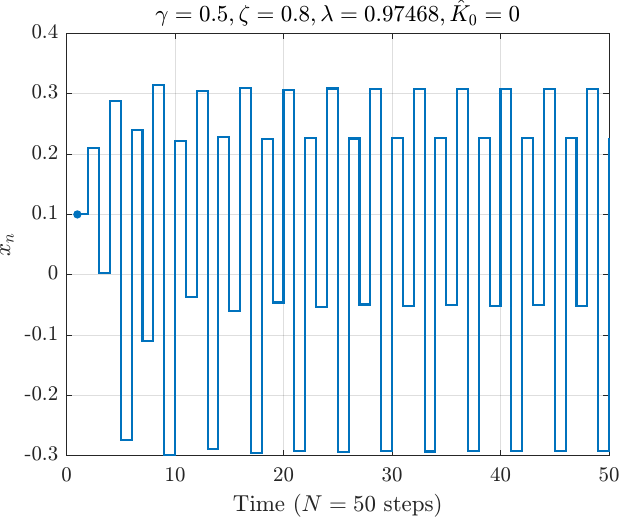}\\
	 	\justify \small Supplementary Figure 1: Overview of the variety of observed regimes. Rows 1, 2 and 3 respectively refer to the regimes R$_2$ (two states), R$_3$ (three states) and R$_4$ (four states i.e.\ period-doubling of R$_2$). 
	 	The left column shows the construction of the iterations on $f$ (thick black line)	using the line $y=x$ (thin black line). The black dot on the x-axis denotes the initial condition $x_0$.
	 	The right column shows the time signal obtained from the iterations on $f$.
	 	The control parameters are specified at the top of each graph.
	 \end{figure}
	  \clearpage
}
\newpage
\section{Results}\label{sec:results}
\subsection{Oscillation threshold}
The region in which the equilibrium R$_1$ is stable is represented in the $(\gamma, \zeta, \hat{K}_0)$ space on Figure \ref{fig:osc}. {Different views and cross-sections are given in the Supplementary Figure 2}. 
When $\hat{K}_0=0$, the well-known oscillation threshold curve is recognized in the $(\gamma, \zeta)$ hyperplane{\cite{dalmont2005analytical, karkar2012oscillation}} {for $\gamma<1$. 
This curve shows the location of Hopf bifurcations, which indicate that the system transitions from equilibrium (no sound) to oscillations (sound).}
{The threshold $\gamma_\mathrm{osc \searrow}$ at which the oscillations start when decreasing the blowing pressure from a reed blocked against the mouthpiece (sometimes named ``inverse threshold''\cite{dalmont_oscillation_2007}) is also visible at $\gamma=1$.}
When $\hat{K}_0$ increases, the oscillation threshold {(noted $\gamma_\mathrm{osc \nearrow}$)} shifts to higher values of $\gamma$ while the inverse threshold stays at $\gamma=1$.
The three cuts represented by dashed lines show that for $\hat{K}_0=\{1,5,10\}$, the average oscillation threshold (computed for $\zeta>0.2$) is respectively at $\gamma\approx \{0.40, 0.60, 0.73\}$.
{For $\hat{K}_0>34$, the stability region of R$_1$ expands along the $\zeta$ axis. In the limit $\hat{K}_0\to +\infty$, equilibrium remains stable throughout the $(\gamma, \zeta)$ plane for $\gamma<1$.}
\begin{figure}[h!]
	\centering
	\includegraphics[width=.6\textwidth]{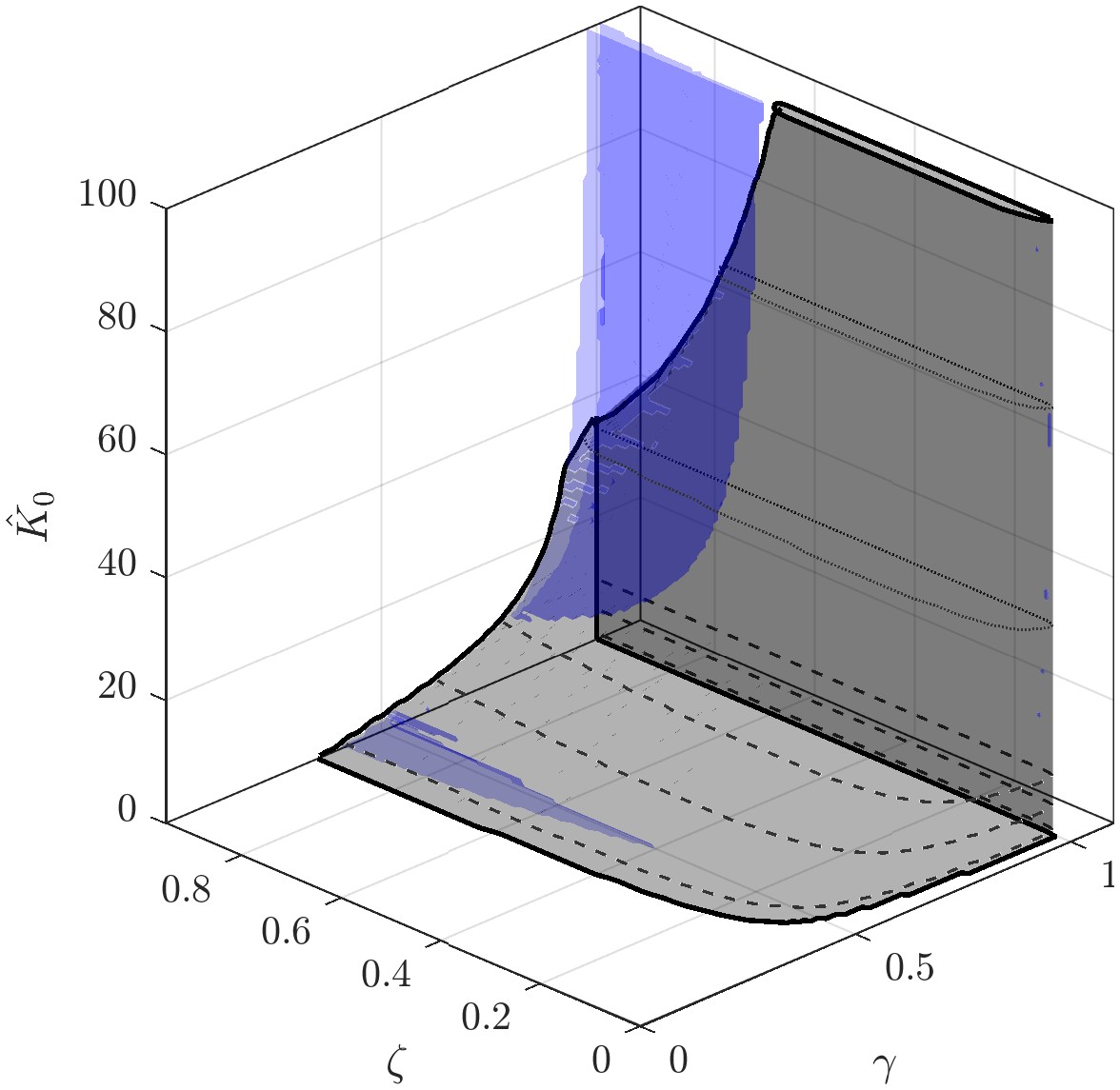}
	\caption{Limit of stability of R$_1$ (in black) and R$_n$ ($n>2$, in blue) in the $(\gamma, \zeta, \hat{K}_0)$ space, for $\lambda=\sqrt{0.95}$.
	{The non-oscillating state (silence) is stable outside the region defined by the black surface. Inside the surface, an oscillating regime must be produced.}
		The three dashed-line curves are horizontal cuts at $\hat{K}_0=\{1, 5, 10\}$.
	The two dotted-line curves are horizontal cuts at $\hat{K}_0=\{35, 70 \}$.	
	The {long-period} regimes are unstable outside the volume delimited by the blue surface.
	{Other views and cross-sections are shown in the Supplementary Figure 2.}
	}
		\label{fig:osc}
\end{figure}
\afterpage{
    \clearpage
    \begin{figure}[p]
	 	\centering
	 	\includegraphics[width=.48\textwidth]{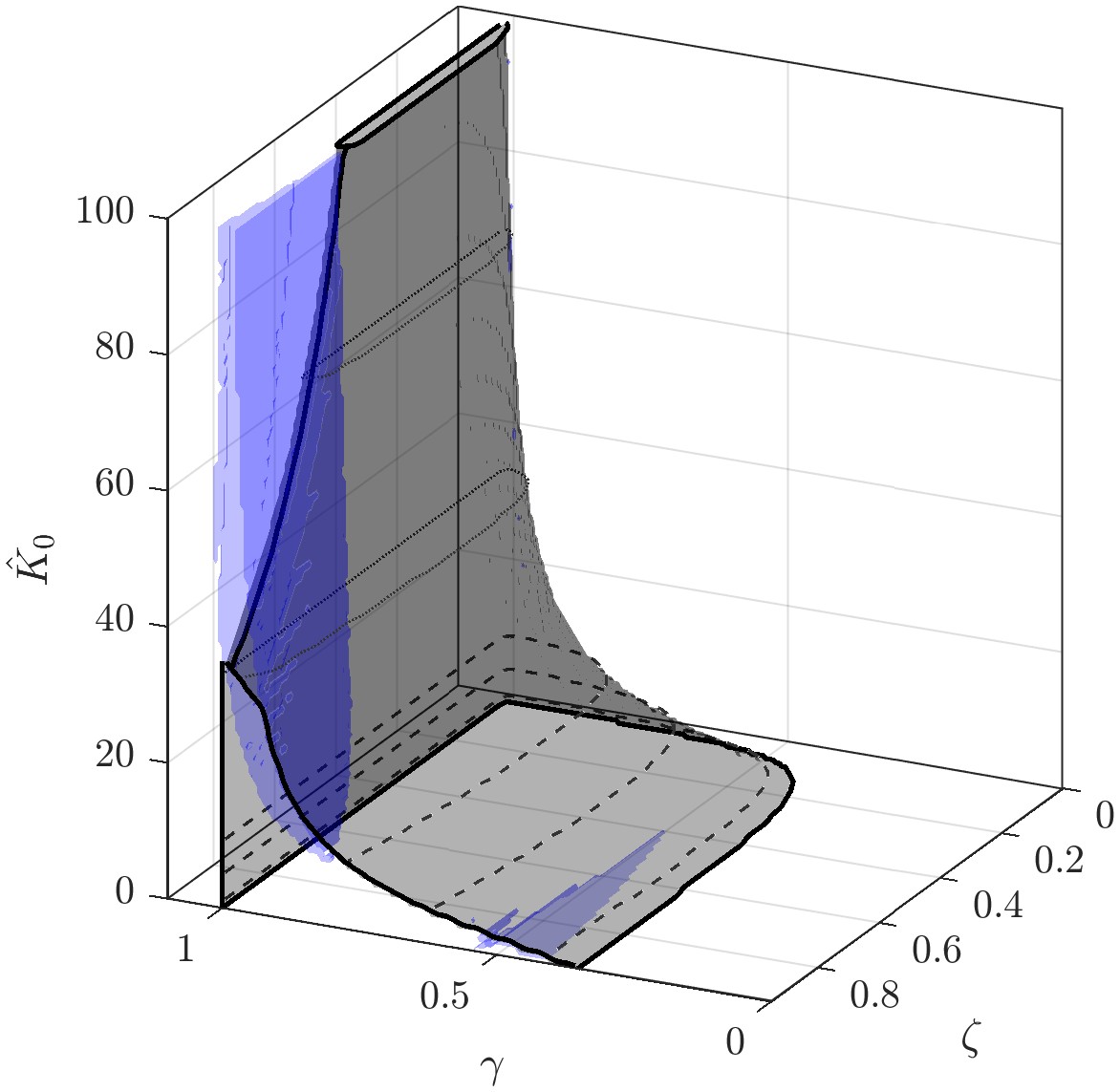}\hfill
	 	\includegraphics[width=.48\textwidth]{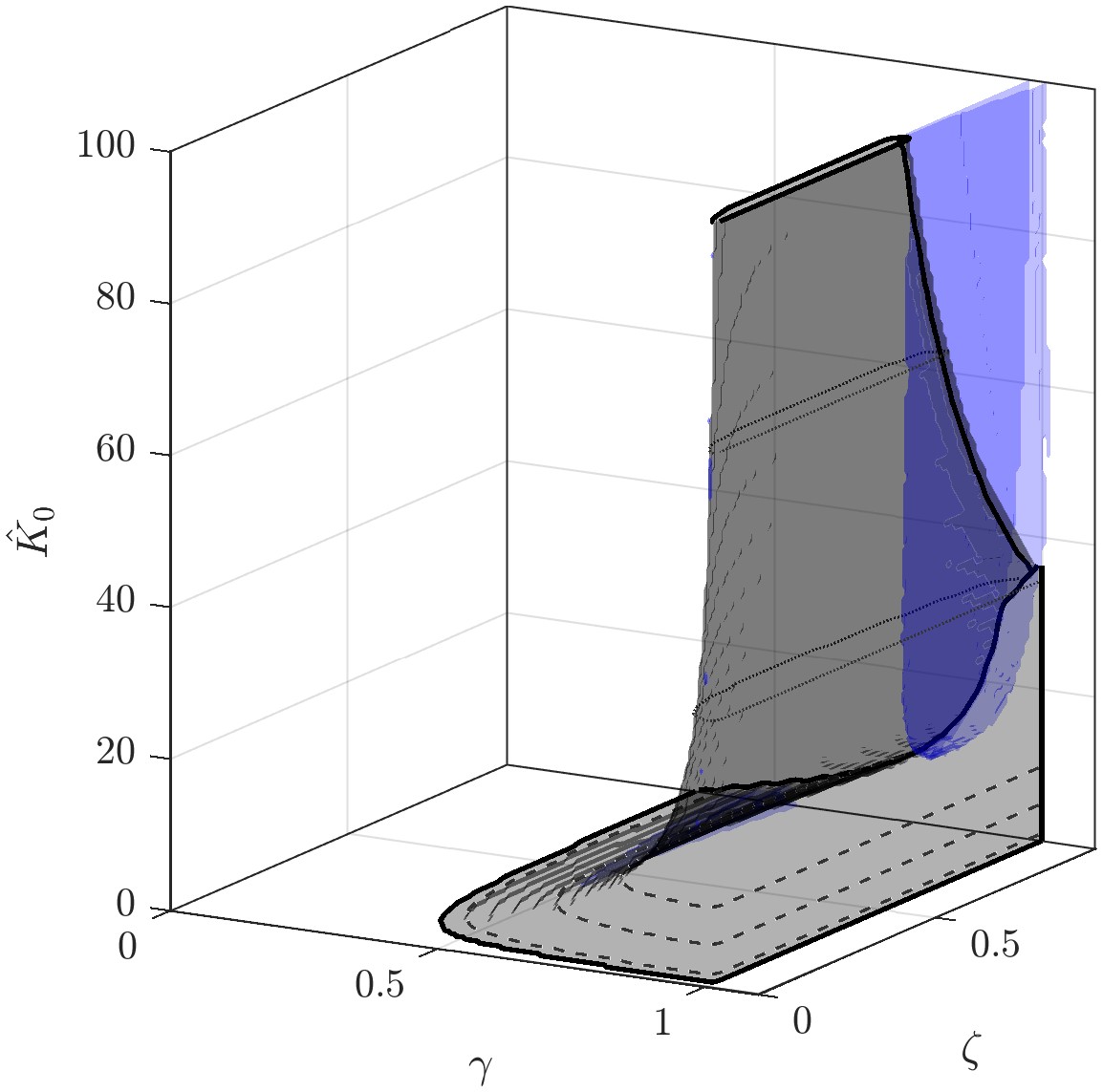}\\
	 	\includegraphics[width=.3\textwidth]{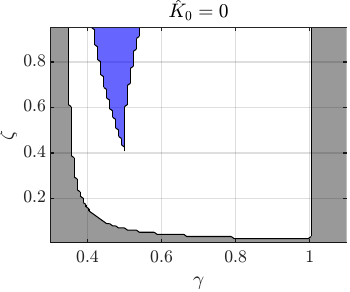}
	 	\includegraphics[width=.3\textwidth]{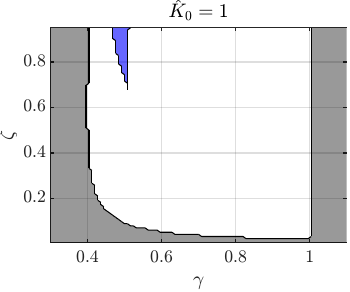}
	 	\includegraphics[width=.3\textwidth]{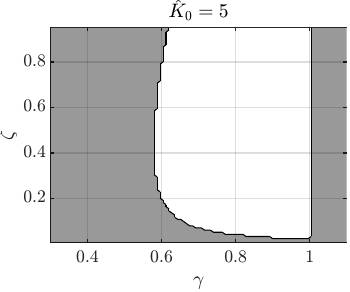}
	 	\includegraphics[width=.3\textwidth]{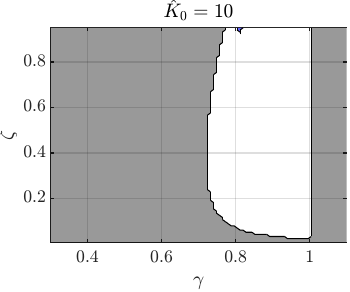}
	 	\includegraphics[width=.3\textwidth]{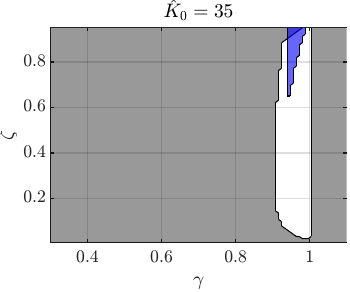}
	 	\includegraphics[width=.3\textwidth]{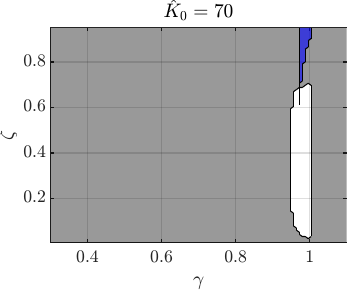}\\
	 	\justify \small	 Supplementary Figure 2: Limit of stability of R$_1$ (in black) and R$_n$ ($n>2$, in blue) in the $(\gamma, \zeta, \hat{K}_0)$ space, for $\lambda=\sqrt{0.95}$.
	On the 3D Figures, The non-oscillating state is stable outside the volume delimited by the black surface, and unstable inside.
	The three dashed-line curves are horizontal cuts at $\hat{K}_0=\{1, 5, 10\}$, also shown on the second and third rows.
	The two dotted-line curves are horizontal cuts at $\hat{K}_0=\{35, 70 \}$, also shown on the third row.	
	The long-period regimes are unstable outside the volume delimited by the blue surface.
	 \end{figure}
    \clearpage
}
\newpage
\subsection{Extinction threshold and {stability of the two-state regime} above $\gamma=1$}
\begin{figure*}[h!]
	\centering
	\includegraphics[height=6cm]{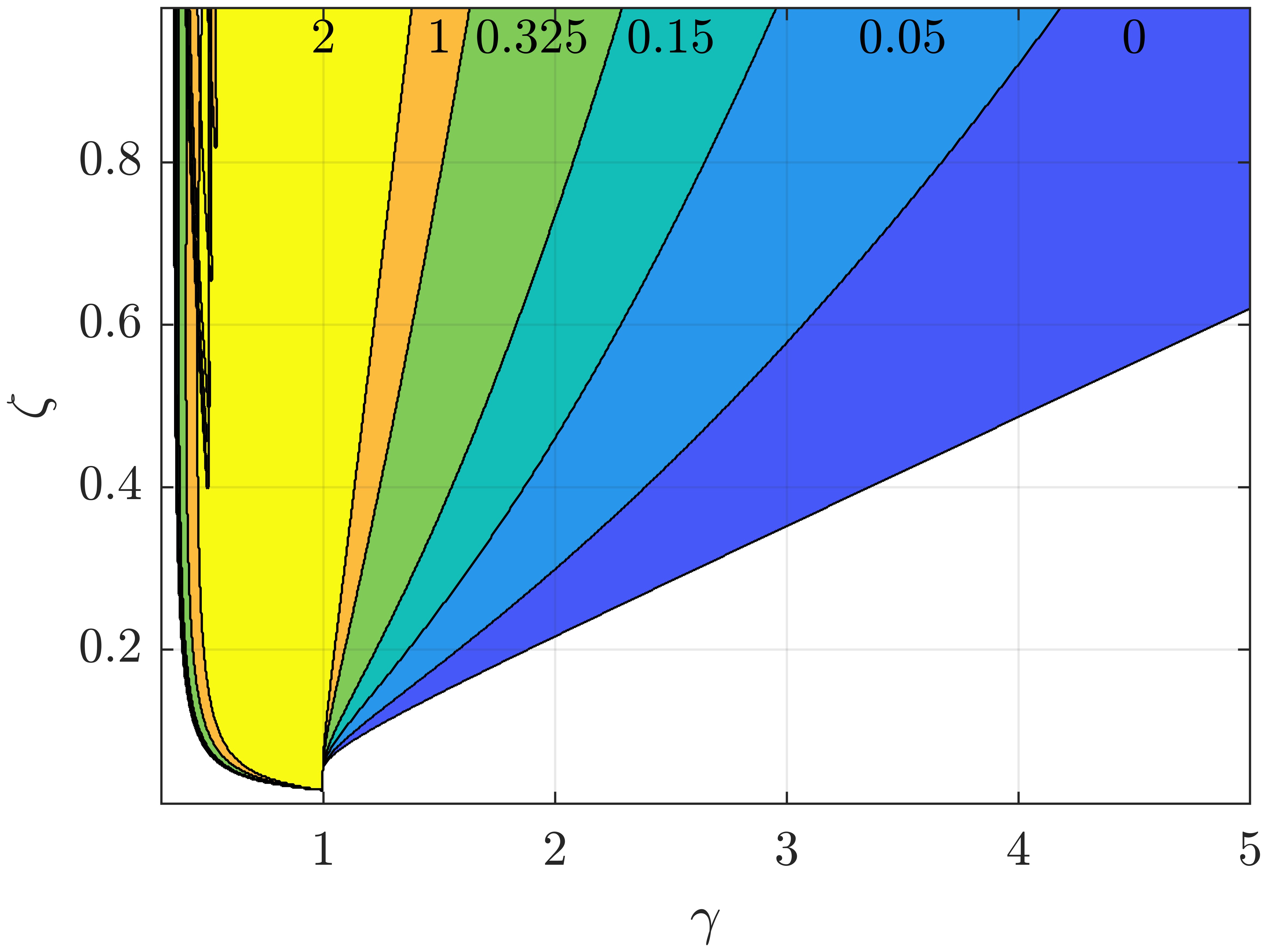}
	\hfill
	\includegraphics[height=6cm]{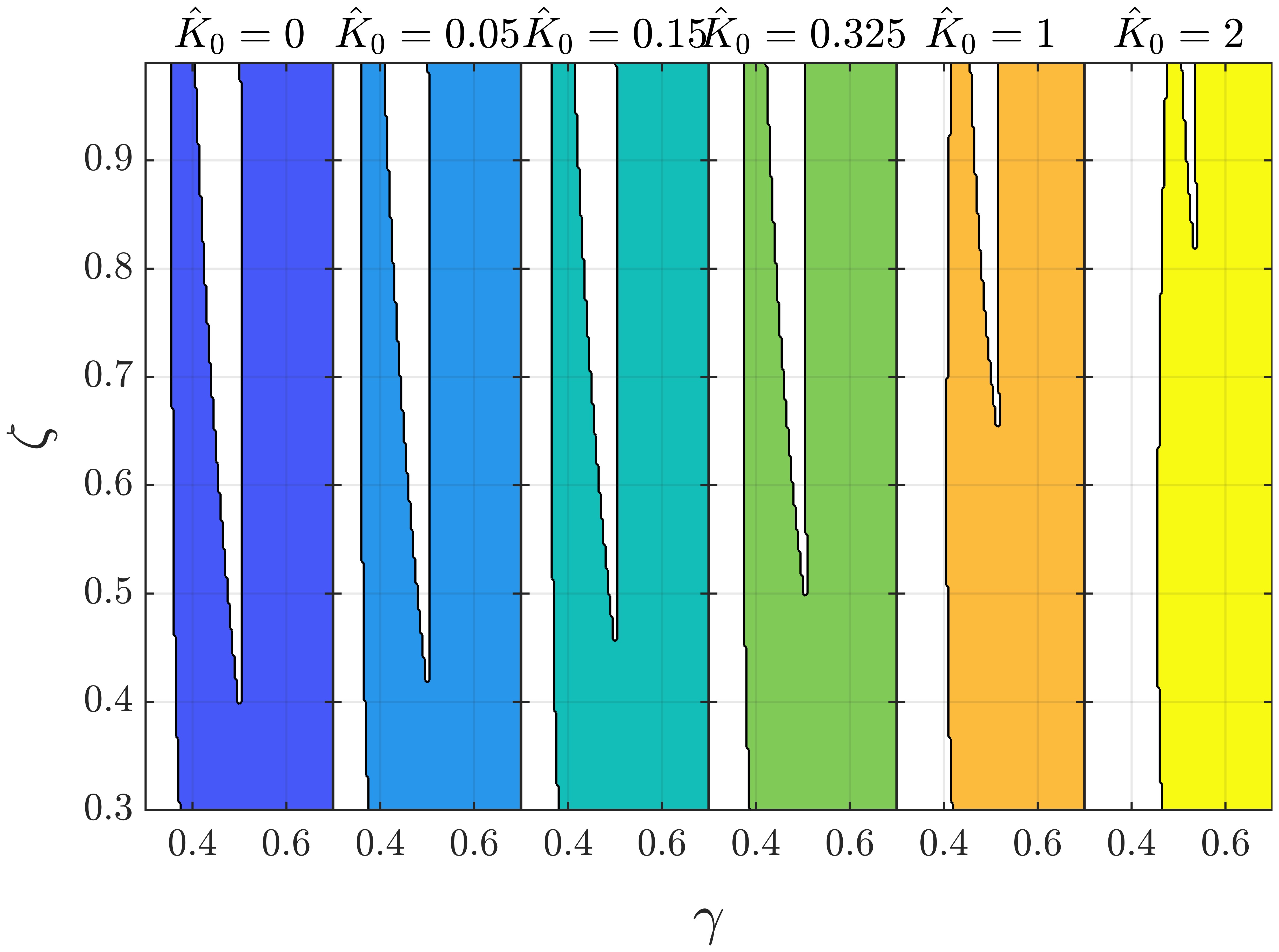}
	\caption{Stability region of R$_2$ in the $(\gamma, \zeta)$ plane, for different values of $\hat{K}_0 \in [0,2]$ and $\lambda=\sqrt{0.95}$. The values of $\hat{K}_0$ corresponding to the different colors are written on the figure.
	{On the left panel,} the color patches are superimposed in a fan shape from blue ($\hat{K}_0=0$) to yellow ($\hat{K}_0=2$).
	{On the right panel, a detailed view is given on the region where R$_2$ is unstable and R$_n$ is stable.}}
	\label{fig:R1}
\end{figure*}

The stability region of {the two-state regime} R$_2$ is represented on Figure \ref{fig:R1}, in the $(\gamma,\zeta)$ plane, for $0\leq\hat{K}_0\leq2$.
{This Figure is also represented in terms of measurable variables (blowing pressure, height of the reed at rest) in Suppl.\ Fig.\ 3.}
Above $\gamma=1$, nonlinear losses greatly reduce the stability of R$_2$, especially for high values of $\zeta$.
This effect is known and has been proven by numerical and experimental comparisons\cite{atig2004saturation,dalmont_oscillation_2007}. 
When $\hat K_0\to \infty$, R$_2$ is unstable everywhere in the $(\gamma, \zeta)$ space.

Furthermore, the effect of localized nonlinear losses on the extinction threshold is much more important than their effect on the oscillation threshold.
For {$\hat{K}_0=\{ 0, 0.325\}$} the  oscillation threshold for $\zeta=0.3$ {(average value for clarinet playing \cite{dalmont_oscillation_2007})} is located respectively at {$\gamma_\mathrm{osc \nearrow}=\{0.378, 0.393\}$, amounting to an increase of 4.0~\%.
In comparison, the value of the extinction threshold of R$_2$ decreases by 40~\% for the same set of control parameters. 
The effect of nonlinear losses is then ten times more important on the extinction threshold than on the oscillation threshold.
This scale discrepancy can explain why the influence of  localized nonlinear losses located at the open end of a tube is not mentioned in previous studies\cite{atig2004termination, dalmont_oscillation_2007, guillemain2005real, szwarcberg2023amplitude, szwarcberg2023FA}.
In addition, in these studies, the difference in the onset of the oscillations might be masked by the bifurcation delay induced by the time-variation of $\gamma$\cite{bergeot2014response, bergeot2013prediction}.}

Finally, the presence of a small region {around $\gamma \approx 0.5$} where the two-state regime is unstable can be noted.
{As shown on the right panel of Figure \ref{fig:R1}, when $\hat K_0$ increases from 0 to 2, the area where R$_2$ is unstable shrinks towards high values of $\zeta$.}
For $\hat{K}_0=0$, a chaotic behavior exists around $\gamma\approx 0.45$ for high values of $\zeta$\cite{taillard2010iterated}.
In the case of $\hat{K}_0=2$, stable four-state and eight-state regimes are found.

\afterpage{
    \clearpage
    \justify \small	Supplementary Table 1: Geometric parameters used in the simulations. $P_M$ is the minimal reed closing pressure, $R$ is the inner radius such that $S=\pi R^2$, $w$ is the width of the reed channel (Dalmont \& Frapp\'e, 2007, {JASA}).
	 It is recalled that $\zeta = w H_0 Z_c \sqrt{2/(\rho_0 P_M)}$ and $\gamma=P_\mathrm{blow}/P_M$.
\begin{table}[h!]
	\centering
	\begin{tabular}{ccc}
	\hline
	$P_M$ [kPa] & $R$ [mm] & $w$ [mm]\\
	8.5 & 7.5 & 12\\	\hline
	\end{tabular}
	\end{table}
	\vspace*{2cm}
	\begin{figure}[h!]
		\centering
		\includegraphics[width=.6\textwidth]{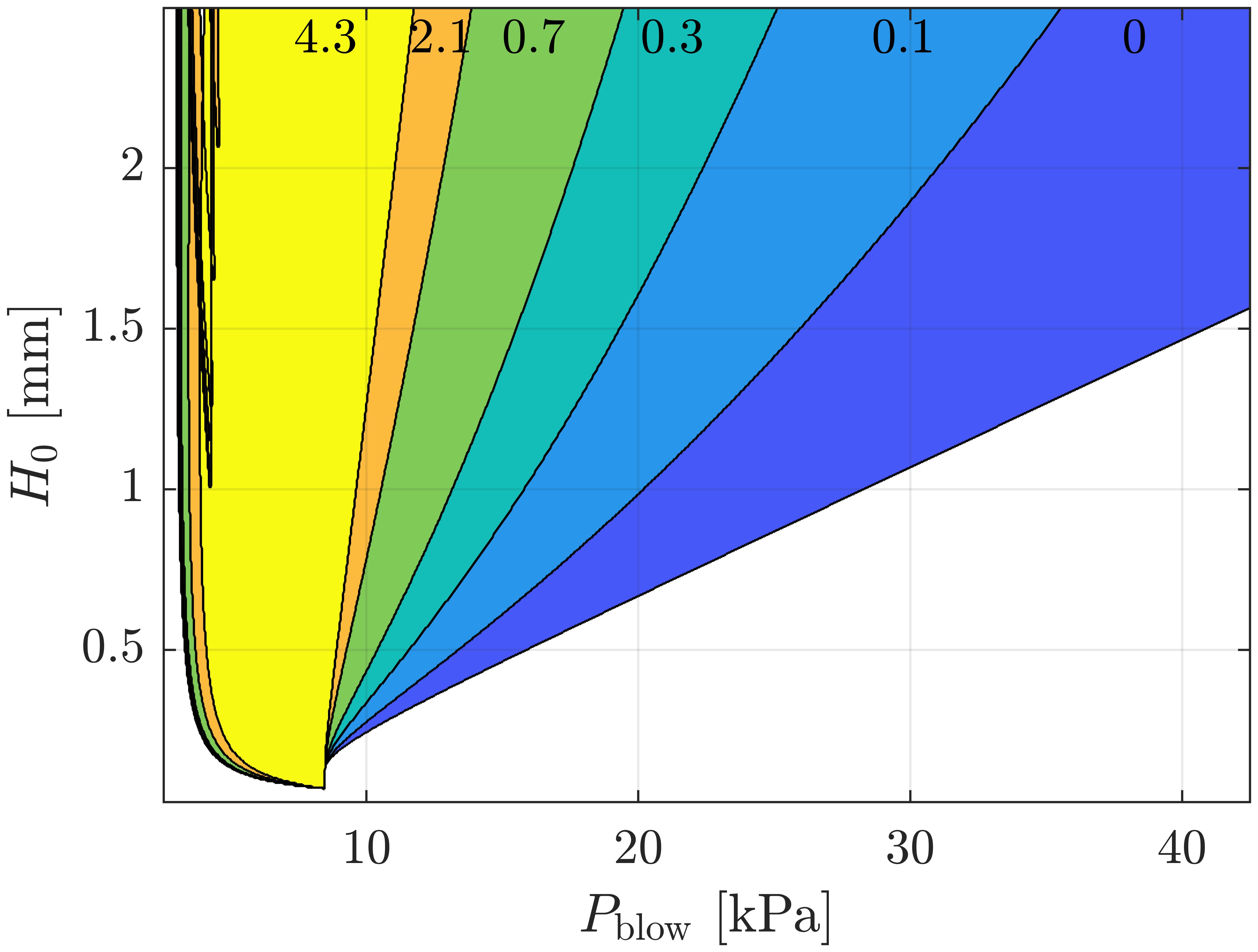}
		\includegraphics[width=.6\textwidth]{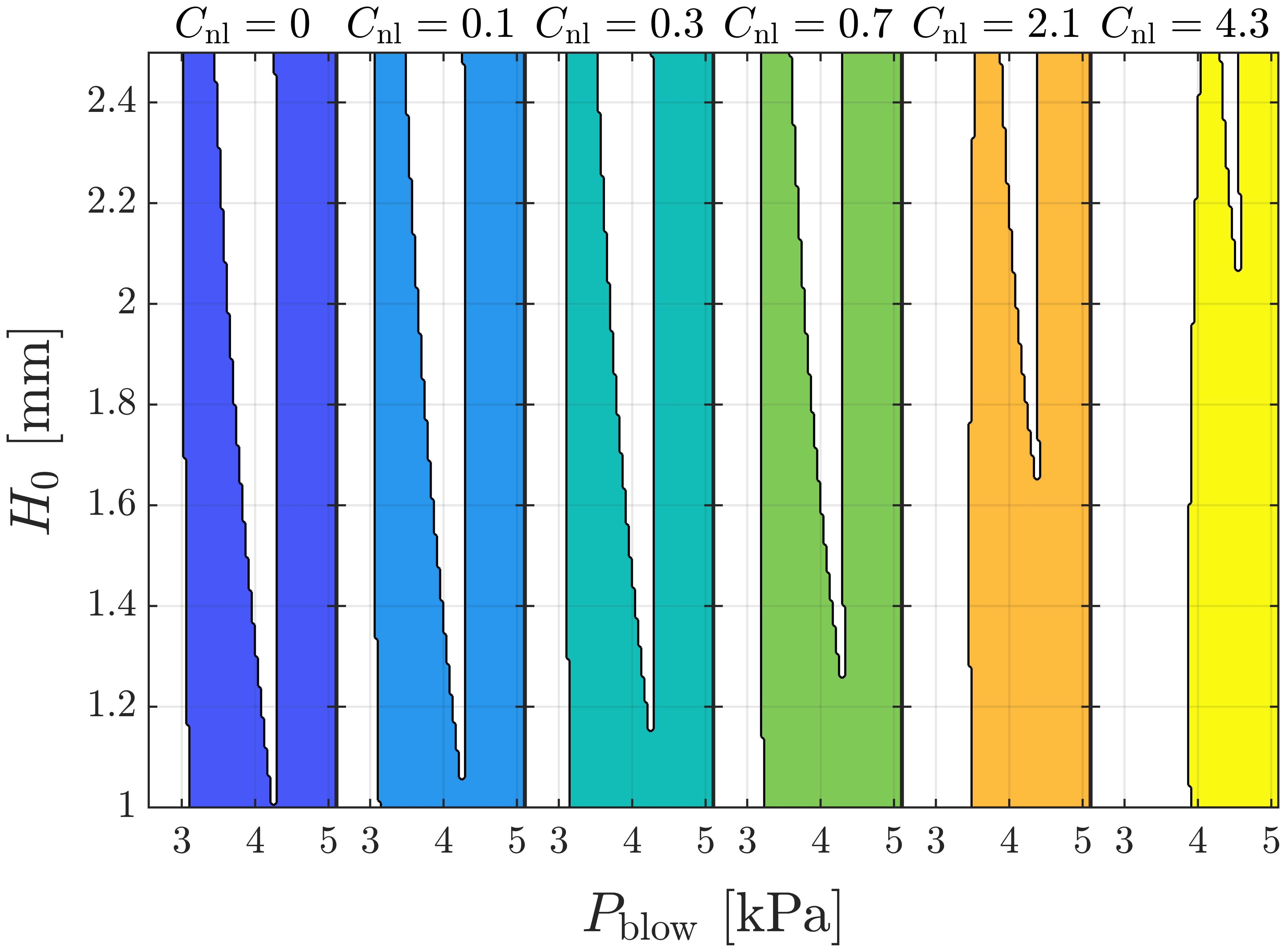}\\
\justify \small	Supplementary Figure 3: Stability region of R$_2$ in the $(P_\mathrm{blow},H_0)$ (blowing pressure, reed height at rest) plane, for different values of $C_\mathrm{nl}\in [0,4.3]$ and $\lambda=\sqrt{0.95}$. 
		Conversion to dimensionless variables can be done using Suppl.\ Table \ref{tab:2}.
		Top panel: values of $C_\mathrm{nl}$ corresponding to the different colors are written on the Figure. The color patches are superimposed in a fan shape from blue ($C_\mathrm{nl}=0$) to yellow ($C_\mathrm{nl}=4.3$). Bottom panel:  detail view on the region where the two-state regime is unstable.
	\end{figure}
	\clearpage
}\newpage
\subsection{{Long-period} regimes}
Regimes that are stable {in either} R$_3$, R$_4$, R$_6$ and R$_8$ are indicated by the volumes delimited by the blue surfaces on Figure~\ref{fig:osc}.
When $\hat{K}_0=0$, stable {long-period} regimes exist around $\gamma=0.5$, {as also shown on Figure \ref{fig:R1}}. 
They progressively disappear when $\hat{K}_0$ increases slightly.
{This reduction may contribute to the fact that period-doubling regimes (R$_4$) are seldom observed on the clarinet.
Note, however, that for $\zeta<0.4$ (a typical value for the clarinet\cite{dalmont_oscillation_2007}) there are no existing long-period regimes for $\hat K_0=0$ and $\lambda=\sqrt{0.95}$, as shown on the detailed view of Figure \ref{fig:R1}.}

From {$\hat{K}_0=9$} and above, {long-period regimes} reappear for very high values of $\zeta$, in a thin band of $\gamma$ near the oscillation threshold.
This band expands towards lower values of $\zeta$ as $\hat{K}_0$ increases.
Above {$\hat{K}_0=34$}, {R$_n$} becomes multistable with the equilibrium R$_1${: for a same set of control parameters, R$_1$ and a long-period regime coexist and are both stable.}
Further computations show that above $\hat{K}_0=100$, this multistability zone expands towards lower values of $\zeta$, as the band of $\gamma$ becomes narrower and closer to $\gamma=1$.
For example, at $\hat{K}_0=500$, for $\gamma=0.993$, the equilibrium is the only existing regime $\forall \zeta <1$.
For $\gamma=0.994$, a stable two-state regime emerges at $\zeta=0.12$, and for $\gamma=0.995$, a stable four-state regime appears at $\zeta=0.64$.
No stable oscillating regime is found for $\hat{K}_0\to \infty$ and $\gamma<1$.
%\newpage
\section{Discussion}\label{sec:discussion}

{To better understand the influence of localized nonlinear losses on the dynamic behavior of a Raman clarinet, a parallel study between linear losses ($\lambda$) and nonlinear losses ($\hat K_\mathrm{0}$) is carried out.}
{The two phenomena are compared at the end of Section \ref{sec:discussion_nl}.
Implications for the musician are discussed in Section \ref{sec:musician}.}
\begin{figure*}[h!]
	\centering
	\includegraphics[width=.7\textwidth]{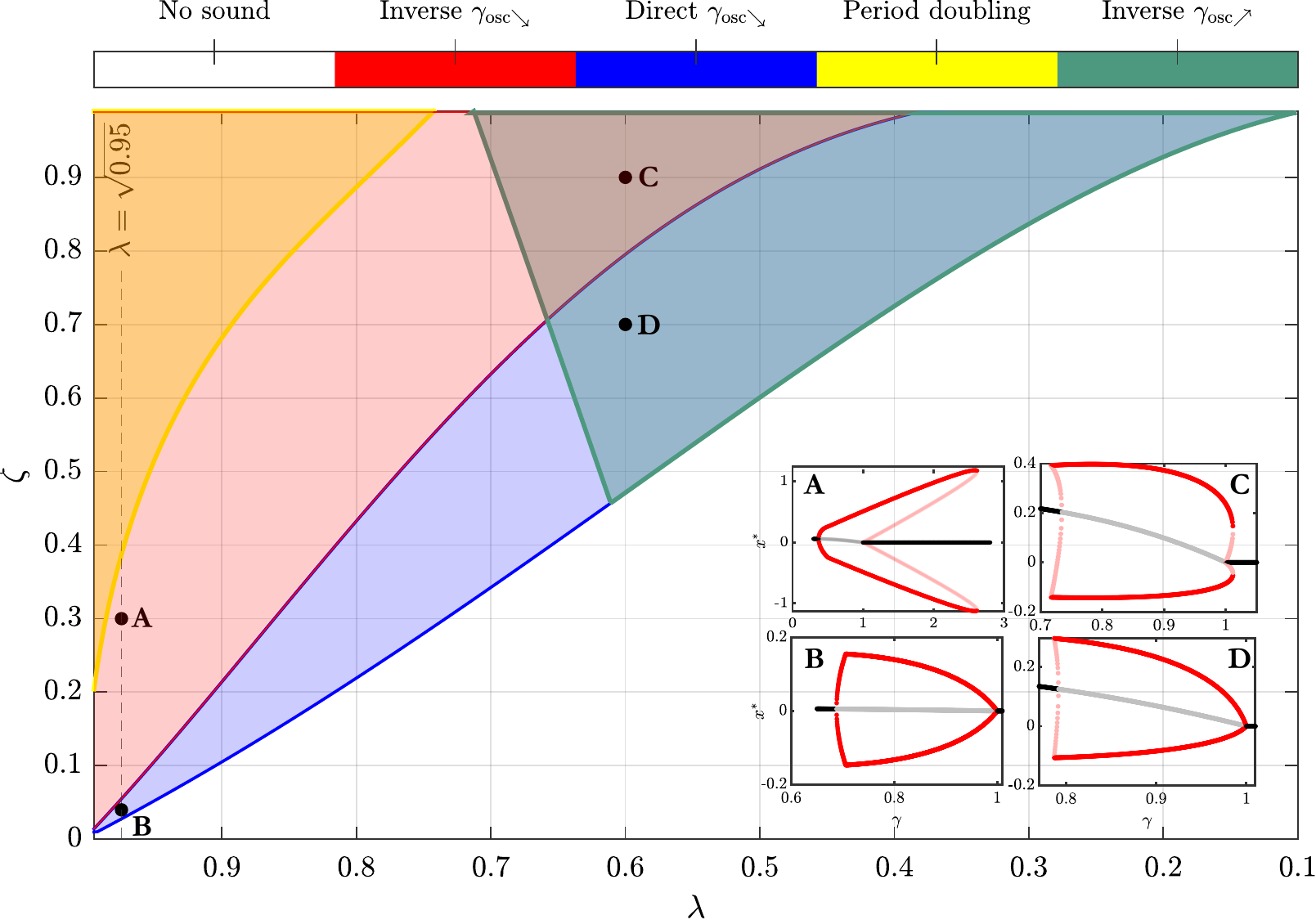}
	\caption{{
	Evolution of the dynamic behavior of the clarinet with respect to the linear losses parameter $\lambda$ and to the embouchure parameter $\zeta$, without localized nonlinear losses.
	The nature of the oscillation thresholds $\gamma_\mathrm{osc\nearrow}$ and $\gamma_\mathrm{osc\searrow}$ is indicated for each point in the $( \lambda, \zeta)$ plane. In the white region, the clarinet produces no sound for {any value of $\gamma$}, whereas sound can be played in the colored regions. 
	In the green region, the bifurcation at $\gamma_\mathrm{osc \nearrow}$ is inverse; outside this region, it is direct.
	In the red region, the bifurcation at $\gamma_\mathrm{osc \searrow}$ is inverse (i.e.\ $\gamma_\mathrm{ext}>1$){, while} in the blue region, it is direct (i.e.\ the clarinet can not be played {for $\gamma>1$}).
	In the yellow region, longer-period regimes R$_n$ ($n\geq 3$) can be played.
	The black dots A, B, C and D {represent} four typical scenarios: {the bifurcation} at $\gamma_\mathrm{osc \nearrow}$ {is direct for A and B and} inverse for C and D, {while }at $\gamma_\mathrm{osc \searrow}$, {it is }direct for B and D {and} inverse for A and C).
	}}
	\label{fig:lambda_zeta}
%\end{figure*}
%\begin{figure*}
	\centering
	\includegraphics[width=.7\textwidth]{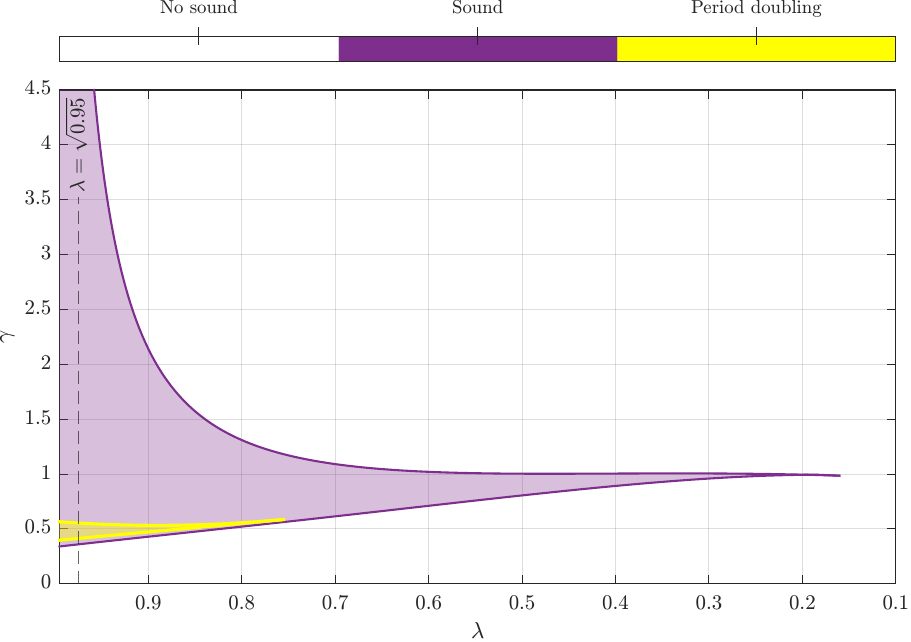}
	\caption{{
	Evolution of the playing range of the clarinet with respect to the linear losses parameter $\lambda$ and  the blowing pressure parameter $\gamma$, for $\hat K_0=0$. {For each $\lambda$, the range of $\gamma$ values allowing an oscillating regime (two-state in purple, longer-period in yellow) is shown, while regions where no sound can be played appear in white.}
	 }}
	 \label{fig:lambda_gamma}
\end{figure*}
\begin{figure*}[h!]
	\centering
	\includegraphics[width=.7\textwidth]{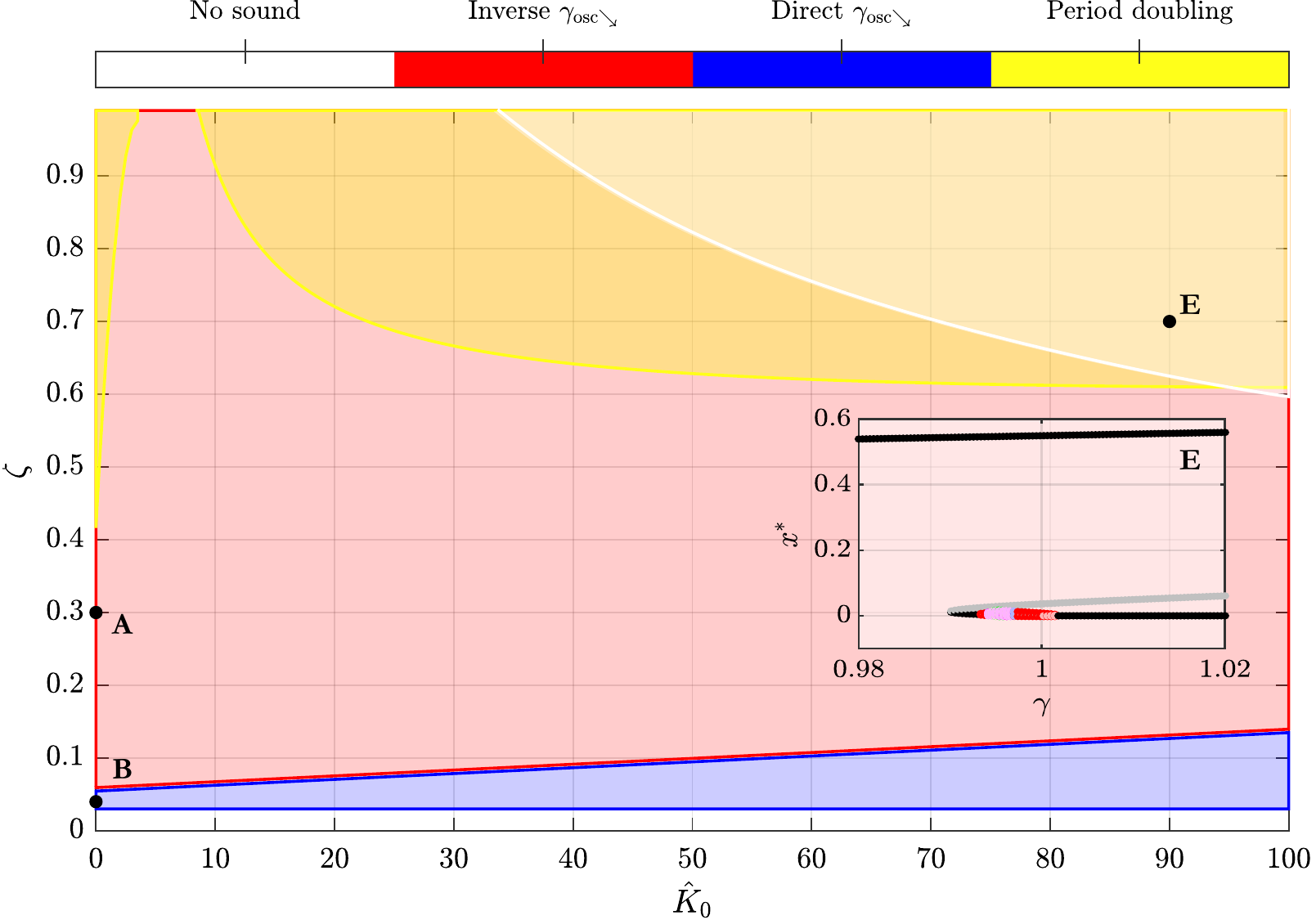}
	\caption{{
	Same as Figure \ref{fig:lambda_zeta}, here in the plane ($\hat K_0, \zeta$) with $\lambda=\sqrt{0.95}$. In the white region at the top right corner, the equilibrium R$_1$ is stable for all $\gamma$, but oscillating regimes are also stable.
	Point E highlights the multistability between the equilibrium (stable in black/unstable in gray) and the oscillating regimes (stable/unstable R$_2$ in red/light red, stable R$_4$ in pink.)
	}}
	\label{fig:K_zeta}
%\end{figure*}
%\begin{figure*}[h!]
	\centering
	\includegraphics[width=.7\textwidth]{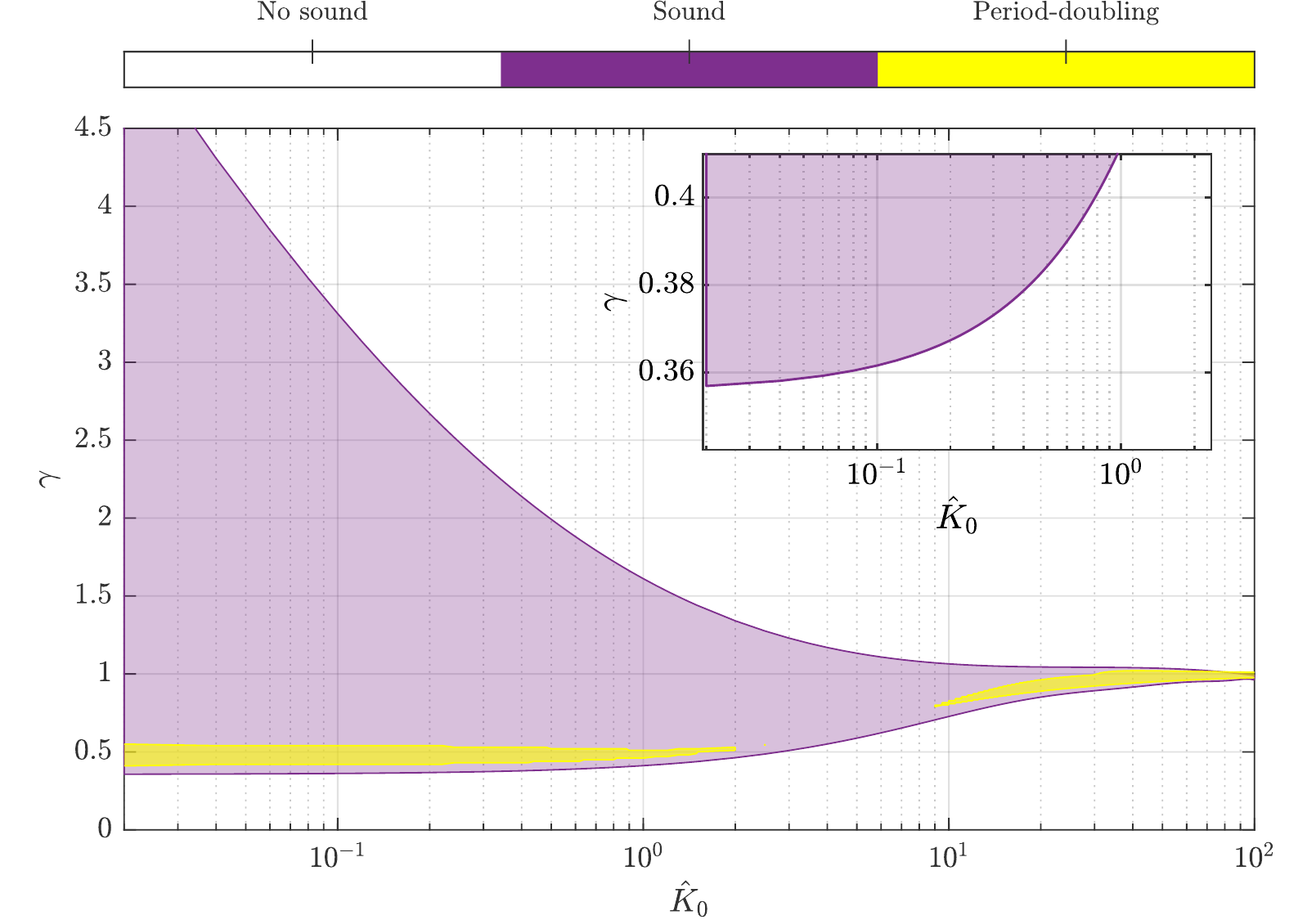}
	\caption{{
	Same description as Figure \ref{fig:lambda_gamma}, in the plane $(\hat K_0, \gamma)$ with $\lambda=\sqrt{0.95}$. Here, the x-axis is in logarithmic scale.}}
	\label{fig:K_gamma}
\end{figure*}
\subsection{{Comparison between linear and localized nonlinear losses}}

\subsubsection{{Linear losses}}\label{sec:discussion_lin}
	{With regard to linear losses, an insightful result is presented in Figure 5 of Taillard and Kergomard (2015)\cite{taillard2015analytical}.
	This Figure {illustrates} the nature of the playing regimes and the type of the bifurcation at the onset of the oscillations, {both} for an increasing blowing pressure ($\gamma_\mathrm{osc\nearrow}$) and {for} a decreasing blowing pressure from $\gamma>1$ ($\gamma_\mathrm{osc\searrow}$).
	This description is {given} in the plane ($\mu, \zeta$), where $\mu=(1-\lambda^2)/(1+\lambda^2)$.
	}
	
	{The results from Taillard and Kergomard (2015)\cite{taillard2015analytical} are reproduced {in} Figure~\ref{fig:lambda_zeta} {using} the method {described} in Section \ref{sec:methods}, for $\hat K_0=0$ and the values of $\gamma, \zeta, \lambda$ written in Table~\ref{tab:1}.
	The different colors in the $(\lambda, \zeta)$ plane {represent} the dynamic behavior of the clarinet ({with} the x-axis  inverted).
	Four points A, B, C, D are {marked in}  Figure \ref{fig:lambda_zeta} to {illustrate} different playing scenarios at $\gamma_\mathrm{osc\nearrow}$ and $\gamma_\mathrm{osc\searrow}$. 
		In the detail {subfigures} at the bottom right corner, the equilibrium {state} R$_1$ is represented in black/gray (stable/unstable{, respectively}), {while} the two-state regime R$_2$ {appears} in red/light red (stable/unstable, {respectively}).}
		
		{
	In the white region in the {lower}-right half of the plane, no sound can be {produced} for any value of $\gamma$.
	For low values of $\lambda$ (points C and D), the green region {indicates} that the bifurcation at $\gamma_\mathrm{osc\nearrow}$ is inverse: the musician can not {transition smoothly} from silence to \textit{pianissimo}, which {may} be considered as a drawback. 	 
	Outside {this} green region, the bifurcation at $\gamma_\mathrm{osc \nearrow}$ is direct{, meaning this continuous transition is possible, as seen for points A and B}.
	The blue and red regions {correspond to different types} of bifurcation at $\gamma_\mathrm{osc \searrow}$. 
	In the red region, the bifurcation is inverse (points A and C). 
	In the blue region, it is direct (points B and D), which also {implies} that no sound can be played for $\gamma>1$. 
	In Section 7.3 of Taillard and Kergomard (2015)\cite{taillard2015analytical}, the authors {suggest} that a direct bifurcation at $\gamma_\mathrm{osc \searrow}$ {may benefit clarinetists, as} it ``conducts to a clean \textit{pianissimo}, with a sound richer in high harmonics'' compared to the sound {near the} threshold at $\gamma_\mathrm{osc \nearrow}$.
	Finally, long-period regimes can {occur} in the yellow region, where $\lambda$ is close to 1.
		}
				
		{Figure \ref{fig:lambda_zeta} provides insights into the clarinet’s dynamic behavior but does not explicitly depict the instrument’s playing range in relation to the blowing pressure $\gamma$. 		
		To provide an alternative perspective on the effects of linear and localized nonlinear losses, playability is represented in the $(\lambda, \gamma)$ plane in Figure \ref{fig:lambda_gamma}.}
		For $\lambda<0.75$, no long-period regimes (in yellow) can be played, as also shown on Figure \ref{fig:lambda_zeta}.
		{Additionnally, for $\lambda\leq0.6$, playing sound for $\gamma>1$ becomes almost impossible. }
		When $\lambda\leq0.39$, Figure \ref{fig:lambda_zeta} {indicates} that the bifurcation at $\gamma_\mathrm{osc\searrow}$ is direct for all $\zeta${, meaning that no sound can be produced for} $\gamma>1$.

\subsubsection{{Localized nonlinear losses}}\label{sec:discussion_nl}
{
Figures \ref{fig:K_zeta} and \ref{fig:K_gamma} reproduce the results of Figures \ref{fig:lambda_zeta} and \ref{fig:lambda_gamma}, now considering nonlinear losses with $\lambda=\sqrt{0.95}$.
}

{
Figure \ref{fig:K_gamma} shows that the oscillation threshold $\gamma_\mathrm{osc \nearrow}$ increases monotonically with $\hat K_0$ for all $\zeta \in [0,0.99]$. In the zoomed-in view, it rises from 0.36 to 0.41 as $\hat K_0$ increases from 0 to 1.
Figure \ref{fig:K_zeta} indicates that for $\hat K_0>34$, equilibrium R$_1$ remains stable with at least one oscillating regime at high $\zeta$ values, as seen in Figure \ref{fig:osc}.
}

{
Regarding the extinction threshold, Figure \ref{fig:K_gamma} shows that nonlinear losses significantly lower its value across all $\zeta \in [0, 0.99]$. For $\hat K_0>10$, it is nearly at $\gamma=1$. Figure \ref{fig:K_zeta} also reveals that the range where $\gamma_\mathrm{osc\searrow}$ is direct (i.e., the extinction threshold is at $\gamma=1$, shown in blue) expands linearly from $\zeta\in[0.03, 0.05]$ at $\hat K_0=0$ to $\zeta\in[0.03, 0.13]$ at $\hat K_0=100$.
}

{
Finally, Figures \ref{fig:K_zeta} and \ref{fig:K_gamma} confirm that stable long-period regimes exist for high $\zeta$ values at all $\hat{K}_0$, except in the range $\hat{K}_0\in[2,8]$.
}

{The comparison between Figures \ref{fig:lambda_zeta} and \ref{fig:lambda_gamma} and Figures \ref{fig:K_zeta} and \ref{fig:K_gamma} highlights the following similarities and differences between linear and nonlinear losses.}
{First, increasing linear losses and nonlinear losses both reduce the playing range of the clarinet by increasing the value of the oscillation threshold $\gamma_\mathrm{osc \nearrow}$ and reducing the value of the extinction threshold.
Second, as the extinction threshold gets close to 1, the range in $\zeta$ where the bifurcation at $\gamma_\mathrm{osc \searrow}$ is direct increases.
Third, increasing linear and nonlinear losses both reduce the stability of the long-period regimes.
For nonlinear losses, this observation is true for $\hat K_0<8$. 
Above this value, stable long-period regimes reappear.
Finally, when linear losses are high ($\lambda<0.61$), the bifurcation at $\gamma_\mathrm{osc \nearrow}$ becomes inverse for all $\zeta$. 
This is not the case for nonlinear losses, where $\gamma_\mathrm{osc \nearrow}$ is always direct.
}

%\newpage
\subsection{{Implications for the musician}}\label{sec:musician}
{
Taillard and Kergomard (2015)\cite{taillard2015analytical} discuss that when linear losses increase, the reduction of the value of the extinction threshold could be useful for the musician to perform a ``clean \textit{pianissimo}''.
Therefore, localized nonlinear losses might also be beneficial by enhancing this effect.
}

\begin{figure}[h!]
	\centering
	\includegraphics[width=.48\textwidth]{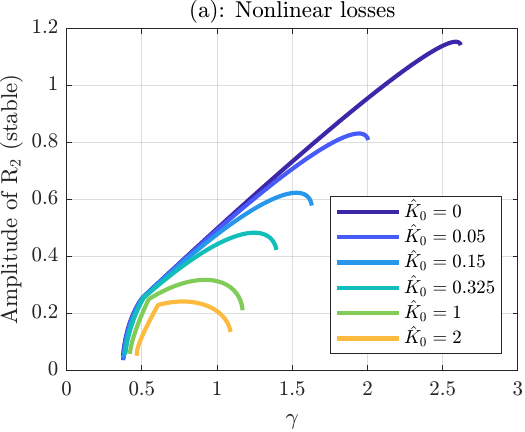}
	\includegraphics[width=.48\textwidth]{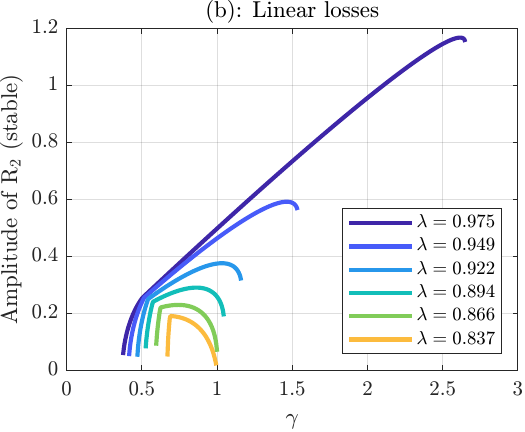}
		\caption{Evolution of the amplitude of R$_2$ (when stable) with respect to the blowing pressure $\gamma$, for $\zeta=0.3$. (a): variation of $\hat K_0$, $\lambda=\sqrt{0.95}$. (b): variation of $\lambda$, $\hat K_0=0$.}
		\label{fig:amp}
\end{figure}

{
Increasing linear and nonlinear losses not only reduces the playing range of the clarinet, but also the amplitude of the acoustic pressure in the air column.
This effect has already been studied experimentally\cite{atig2004saturation, dalmont_oscillation_2007} and numerically \cite{atig2004saturation, guillemain2005real,szwarcberg2023amplitude}.
In comparison with linear losses, it is worth noting that localized nonlinear losses significantly reduce the amplitude of the acoustic pressure without altering too much the value of $\gamma_\mathrm{osc\nearrow}$ nor changing the nature of the bifurcation, as shown on Figure \ref{fig:amp}.
{As $\hat K_0$ increases (Fig.\ \ref{fig:amp}(a)), the slope of the amplitude around $\gamma_\mathrm{osc \nearrow}$ decreases, whereas linear losses increase it (Fig.\ \ref{fig:amp}(b)), eventually leading to an inverse bifurcation (green zone in Figure \ref{fig:lambda_zeta}).}
Thanks to localized nonlinear losses, the reduction of the slope near $\gamma_\mathrm{osc \nearrow}$ could help clarinetists to control the softest dynamics of their instrument.
}

\section{Conclusion}
{
	The main aim of this study is to contribute to the understanding of nonlinear losses that are localized at the open end of a clarinet-like instrument.
			The complete system is written as an iterated function, which enables to investigate the stability of the different playing regimes in a simple and systematic way.
			This work thus follows on from the literature devoted to minimal clarinet models\cite{taillard2010iterated, bergeot2013prediction, taillard2015analytical}.
As a main result of this study, localized nonlinear losses increase the value of the minimal blowing pressure for which the oscillations start.
			This effect is not reported in previous works dedicated to this phenomenon \cite{atig2004saturation, guillemain2005real, dalmont_oscillation_2007}.
			Although the bore of a real clarinet is much more complicated than a cylindrical tube, this result provides some insight for the wind instrument makers into the value of undercutting the side holes to modify the dynamic range of the instrument. 
			This hypothesis could be explored in a future work thanks to the formulation proposed in Eq.\ \eqref{eq:ref0}, which provides a simple way to take into account localized nonlinear losses in the side hole of a waveguide model.}
			{Furthermore, localized nonlinear losses act as a natural compressor, reducing the amplitude of the acoustic pressure without altering the direct nature of the bifurcation at the onset of the oscillations. 
			This is not the case for linear losses, which significantly increase the value of the minimal blowing pressure to produce a sound, and  can change the nature of the bifurcation at the onset of the oscillations from direct to inverse.
			Finally,} the increase of nonlinear losses entails the disappearance of high order regimes, before they reappear for very strong values of the coefficient.
Future studies may explore experimentally the possible influence of localized nonlinear losses {on the control of the nuances from the clarinetist}.

\section*{Acknowledgments}
This study has been supported by the French ANR LabCom LIAMFI (ANR-16-LCV2-007-01). 
The authors warmly thank F. Monteghetti for fruitful preliminary discussions.

\section*{Author declarations and Data availability statement}
The authors have no conflicts to disclose.
The data that support the findings of the present study are available from the corresponding author upon reasonable request.
\footnotesize
\bibliography{biblio}

\end{document}